\documentclass[a4paper]{jpconf}
\usepackage{graphicx}
\usepackage{wrapfig}
\usepackage[small]{caption}
\usepackage{color}

\begin{document}
\title{The software of the ATLAS beam pick-up based LHC monitoring system}

\author{C~Ohm$^1$ and T~Pauly$^2$}

\address{$^1$Department of Physics, Stockholm University, 106 09 Stockholm, Sweden}
\address{$^2$European Organization for Nuclear Research, 1211 Gen\`eve, Switzerland}

\ead{christian.ohm@cern.ch, thilo.pauly@cern.ch}

\begin{abstract}
The ATLAS BPTX stations are comprised of electrostatic button pick-up detectors, located 175 m away along the beam pipe on both sides of ATLAS. The pick-ups are installed as a part of the LHC beam instrumentation and used by ATLAS for timing purposes. The signals from the ATLAS BPTX detectors are used both in the trigger system and for a stand-alone monitoring system for the LHC beams and timing signals. The monitoring software measures the phase between collisions and clock with high accuracy in order to guarantee a stable phase relationship for optimal signal sampling in the sub-detector front-end electronics. It also measures the properties of the individual bunches and the structure of the beams. In this paper, the BPTX monitoring software is described, its algorithms explained and a few example monitoring displays shown. In addition, results from the monitoring system during the first period of single beam running in September 2008 are presented. 
\end{abstract}

\section{Introduction}
\label{sec:introduction}
The ATLAS experiment \cite{detectorpaper} at the Large Hadron Collider (LHC) \cite{lhcmachinepaper} must be synchronized to the proton-proton collisions to ensure the quality of the event data recorded by its sub-detectors. In order to facilitate the synchronization, ATLAS receives timing signals from the LHC machine and has two beam pick-up detectors at its disposal, installed upstream from the interaction point. The signals from these so-called \emph{BPTX} detectors serve two purposes in ATLAS: 
\begin{itemize}
	\item The BPTX signals are used in a stand-alone monitoring system for the LHC beams and timing signals. The system monitors the phase between the collisions and the LHC clock signals that drive the ATLAS electronics in order to discover potential drifts. This system also measures the structure and uniformity of the LHC beams, and properties of their individual bunches.
	\item By discriminating the signals from the BPTX detectors and compensating for the lengths of the transmission lines, the BPTX system provides Level-1 trigger input signals synchronous to bunches passing through ATLAS. These trigger signals serve as an absolute time reference and are particularly useful when timing in the triggers based on real physics signals.
\end{itemize}
A concise description of the ATLAS BPTX system can be found in \cite{bptxnim} and the concepts are described in more detail in \cite{timing-in,masterohm}. This paper describes the software of the beam pick-up based monitoring system for the LHC beams and timing signals, and presents a few results from the first period of single beam running in September 2008.

\subsection{LHC timing signals}

\begin{wrapfigure}{r}{0.35\textwidth}
	\vspace{-20pt}
	\begin{center}
	\includegraphics[width=0.3\textwidth]{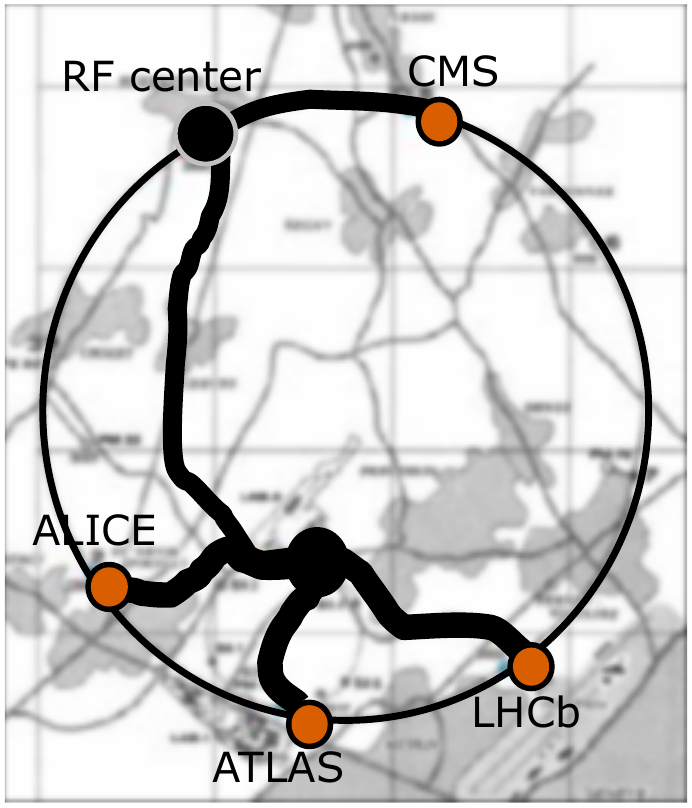}
	\caption{An illustration of how the LHC timing signals are transmitted from the RF center to the four major LHC experiments.}
	\label{fig:timingsignaldistribution}
	\end{center}
\end{wrapfigure}

In order to allow the LHC experiments to be synchronized to the collisions, the accelerator provides them with timing signals related to its beams, transmitted through several kilometers of optical fibers\cite{ttc}. Temperature changes that affect the properties of the fibers or other unforeseen problems with the transmission can introduce phase shifts in the clock signals and cause the ATLAS on-detector electronics to sample the detector signals at a non-optimal working point. In turn, this will jeopardize the quality of the data for the recorded events and the efficiency of the trigger system. Figure~\ref{fig:timingsignaldistribution} shows how the timing signals are distributed from the RF center of the LHC machine to the experiments.

The LHC provides one reference clock signal, \emph{BCref}, corresponding to the frequency at maximum beam energy and two clock signals that are synchronous to the accelerating beams, \emph{BC1} and \emph{BC2}. In addition, two orbit signals \emph{Orbit1} and \emph{Orbit2} with $f_{\mathrm{Orbit}i} = f_{\mathrm{BC}i}/3564$ are provided to mark every LHC turn, synchronized to their respective beams.

\subsection{The ATLAS BPTX detectors}
On both sides of ATLAS, 175\,m upstream from the interaction point, beam pick-up detectors are installed along the LHC beam pipe. These so-called \emph{BPTX stations} are beam position monitors provided by the LHC machine, but operated by the experiments for timing purposes. They are comprised of four electrostatic button pick-up detectors each, arranged symmetrically in the transverse plane around the beam pipe.

\begin{figure}[!!h]
	\centering
	\includegraphics[width=0.25\paperwidth]{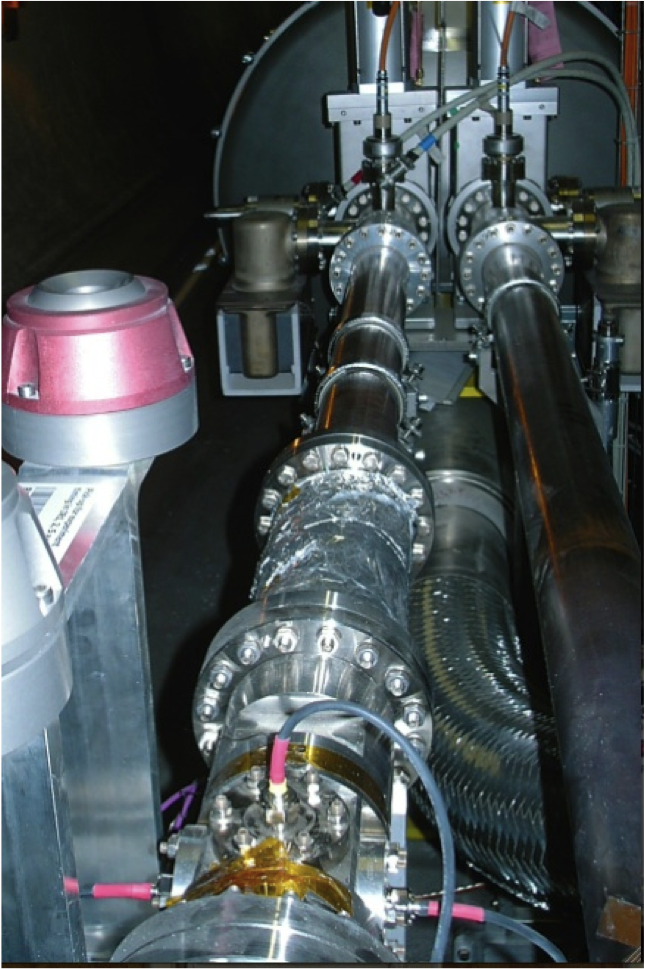}
	\caption{A photograph of one of the two ATLAS BPTX stations.}
	\label{fig:bptxstation}
\end{figure}

Figure~\ref{fig:bptxstation} shows the ATLAS BPTX station for beam 2, installed in the accelerator tunnel on the C-side of ATLAS. At the bottom of the photograph, the cables from three of the four button pick-ups are visible. The signals are combined and transmitted to the underground counting room \emph{USA15} via a $\sim$220\,m low-loss cable. 

\subsection{Usage of the BPTX signals in ATLAS}
\begin{figure}[!!htp]
	\centering
	\includegraphics[width=0.5\paperwidth]{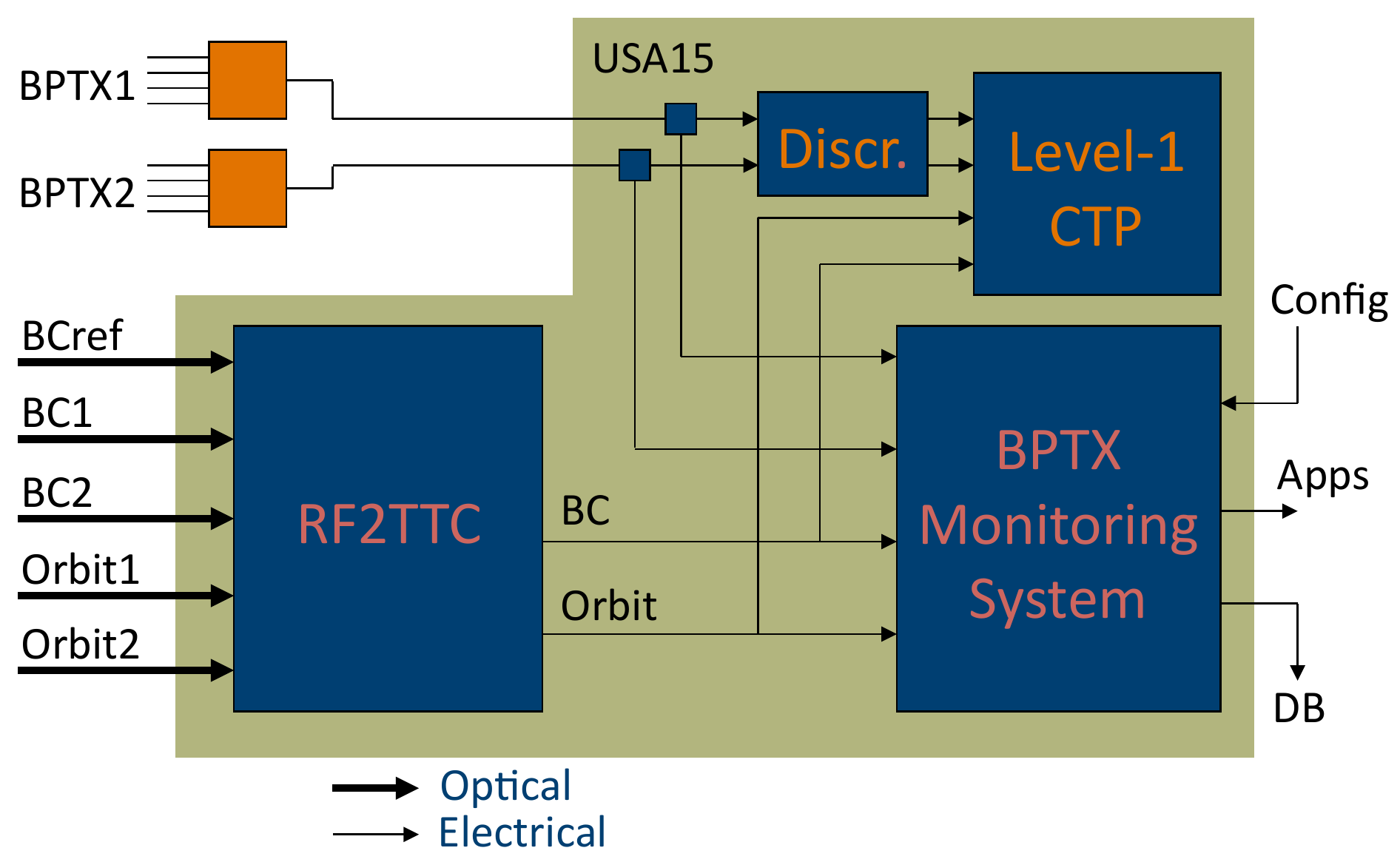}
	\caption{The usage of the BPTX signals in ATLAS.}
	\label{fig:bptxcontext}
\end{figure}

As briefly mentioned in the Introduction, the ATLAS BPTX system serves two purposes. Figure~\ref{fig:bptxcontext} shows how the BPTX signals are used both in the trigger system and for the BPTX monitoring system which will be described in more detail in the following sections. The signals from the four pick-ups are combined to eliminate potential effects of the beam position in the transverse plane\footnote{To first order, the signal amplitude from an individual pick-up is proportional to its closest distance to the passing charge. Since the BPTX stations are installed for timing purposes, the signals from all four pick-ups are summed which effectively cancels out any offset along an axis in the transverse plane.} before they are transmitted to the underground counting room \emph{USA15}. The signals are split and fed into a constant-fraction discriminator to form a beam-related Level-1 trigger input, and into the BPTX monitoring system. The LHC timing signals enter the counting room as optical signals and are converted by the \verb=RF2TTC= timing receiver. The \verb=RF2TTC= module controls what signals are passed through to ATLAS and allows the manipulation of certain parameters, e.g. their phase, pulse duration and polarity.

\section{Overview of the BPTX monitoring system software}
For the monitoring of the LHC beams and timing signals in ATLAS, a flexible software solution was chosen.
The BPTX and LHC timing signals are captured and digitized by a deep-memory, high sampling rate oscilloscope\footnote{LeCroy WaveRunner 64 Xi. This oscilloscope has an analog bandwidth of 600 MHz and a sampling rate of 5 GS/s.}. By default, 100\,$\mu$s are captured, digitized and transferred to a rack-mounted Linux computer installed in \emph{USA15}. Since it takes 89\,$\mu$s for a bunch to travel around the LHC, this guarantees capturing all bunches in the accelerator at least once per acquisition, allowing measurements on a bunch-by-bunch basis. By frequently ($\sim$0.5\,Hz) analyzing these signals together, phase drifts or other problems in the beams or timing signals will be discovered. The analysis is done completely in software and allows measuring and monitoring
\begin{itemize}
	\item the average phase between the timing signals and the collisions
	\item the beam structure and detection of out-of-time bunches, including in the abort gap
	\item individual bunch properties, e.g. intensity, phase and longitudinal length
\end{itemize}

\begin{figure}[!!h]
	\centering
	\includegraphics[width=0.5\paperwidth]{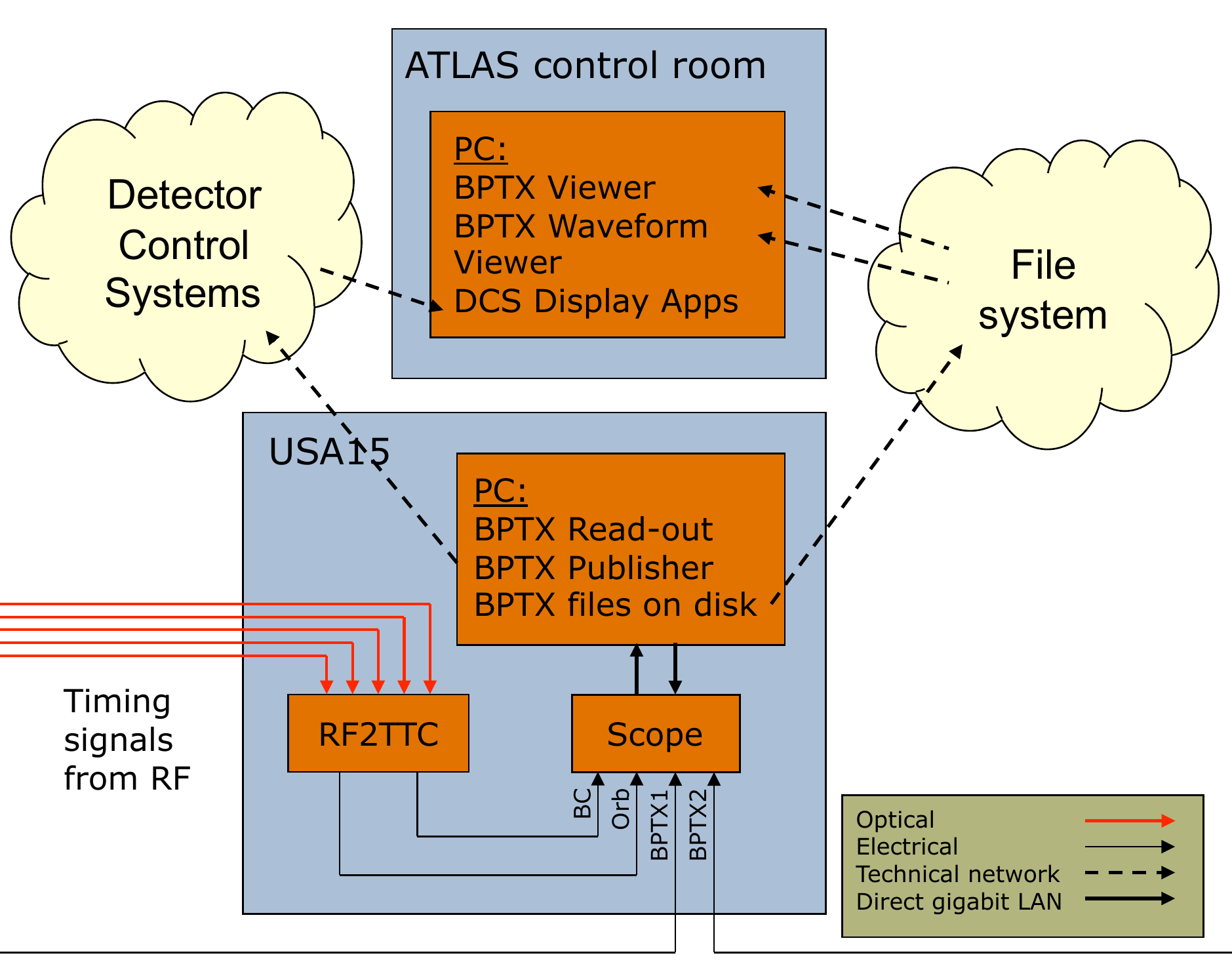}
	\caption{Diagram showing the components of the BPTX monitoring system and how they interact.}
	\label{fig:bptxoverview}
\end{figure}
The software of the BPTX monitoring system is divided into several applications (see Figure~\ref{fig:bptxoverview}) that perform separate, well-defined tasks:
\begin{itemize}
	\item \emph{BPTX Read-out} is the heart of the monitoring software, processes the digitized waveforms, and extracts parameters describing the identified bunches and clock edges. It then combines the condensed information to assign a phase and \emph{Bunch Crossing Identifier} (BCID) to each bunch. The output is saved on disk where it can be read by other components in the monitoring system.
	\item \emph{BPTX Viewer} gives the control room shifter an overview of the LHC beams and timing signals, and offers the possibility to track changes over time and identify oddly behaving bunches.
	\item \emph{BPTX Waveform Viewer} lets the control room shifter inspect saved waveforms to investigate.
	\item \emph{BPTX Publisher} communicates with other parts of the ATLAS software framework and transmits summary data to make it available to other online services and permanently saved in the ATLAS conditions database.
\end{itemize}
The three first components are described in the following sections. For information about the file formats, implementation details and the mathematical model of the signal from the BPTX detectors, the reader is referred to \cite{masterohm}.

\section{BPTX Read-out: signal processing and analysis}
\label{sec:bptxreadout}

\subsection{Signal processing}
This section describes the algorithms that extract the locations of the timing signal pulses and the parameters that describe the LHC bunches. The information extracted from the waveforms are saved in a condensed, lightweight representation called \emph{waveform descriptors}, which are used internally for the rest of the processing, written to disk and read by other applications of the monitoring system. By combining the descriptors for the BPTX and timing signals, each bunch can be assigned a BCID and a phase.

\subsubsection{Clock and orbit signal processing}
Clock signals coming from the RF2TTC module during ATLAS data taking exhibit a rise-time of about 1\,ns which agrees with test measurements \cite{ttc}. With a sampling rate of 5\,GS/s, the edge in the digitized signal will therefore consist of about 5 samples, which allows performing a fit in order to determine the threshold crossing with high precision. During the design phase of the BPTX monitoring system, three methods were considered for determining the exact locations of the clock edges. 
\begin{enumerate}
	\item Linear interpolation between the two sample points around the threshold crossing
	\item Linear fit based on 5 samples points around the threshold crossing
	\item Third order polynomial fit based on 5 samples points around the threshold crossing
\end{enumerate}





All three methods were tested on simulated clock signals exhibiting noise and cycle-to-cycle jitter characteristic of the output signals of the \verb=RF2TTC= module. All three algorithms had sub-picosecond precision. The linear interpolation is used as the default method as it is the computationally least demanding.

\subsubsection{BPTX signal processing}
An LHC bunch can be described by three parameters, its arrival time, its intensity (number of protons) and its longitudinal length. During the development phase of the BPTX monitoring system, several ways of measuring these parameters were considered and tested with simulations.

\begin{figure}[!!htp]
	\centering
	\includegraphics[width=0.7\paperwidth]{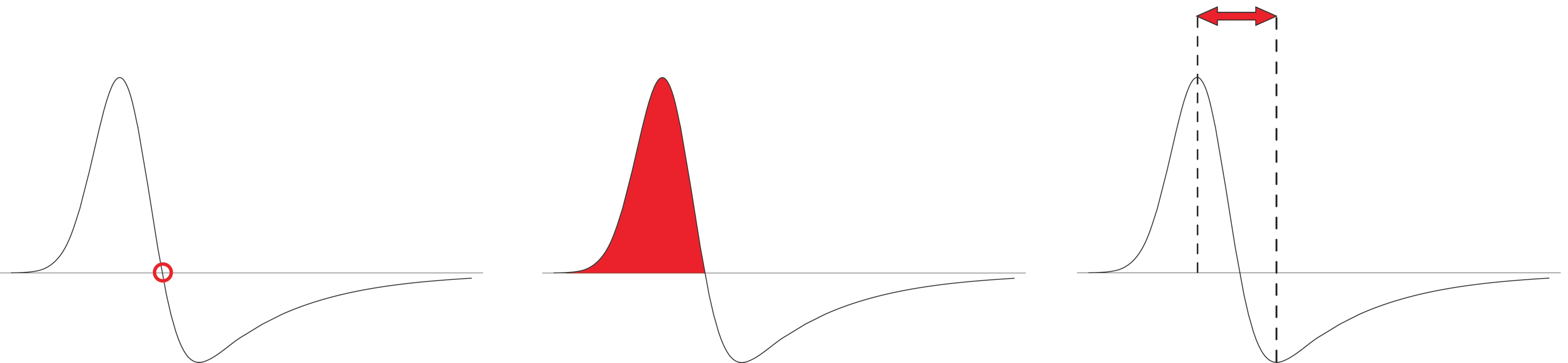}
	\caption{The waveform features measured to determine the bunch arrival time (left) and intensity (middle). The distance between the peak and the valley of the bipolar pulse correlates well with the longitudinal length of the bunch and is therefore used to measure this parameter.}
	\label{fig:bunchparametermeasures}
\end{figure}

When a bunch passes through a BPTX station, it gives rise to a fast bipolar pulse with a frequency spectrum peaking at around 400 MHz and tails reaching a few GHz. The long transmission line and the limited analog bandwidth of the oscilloscope attenuate the high-frequency content and shape the signal accordingly. After quantifying the couplings between different properties of the bipolar pulse and the bunch parameters in a simulation study, the waveform features in Figure \ref{fig:bunchparametermeasures} were chosen. 

The zero-crossing is a sharp vertical edge and thus a well-suited region to pick off the time of the pulse. Moreover, it is stationary with varying bunch intensity. The location of the zero-crossing is determined by solving where the function resulting from a linear interpolation between the last positive and first negative samples crosses zero. Simulations suggest that the location of the zero-crossing relative to the actual arrival time depends slightly on the bunch length, which degrades the resolution of the arrival time measurement if the beams contain bunches of varying length. However, as soon as the mathematical model that the simulations rely upon has been tuned to data, this correlation can be determined and compensated for at the signal processing stage.

The value of the integral under the peak of the pulse is proportional to the number of protons in the bunch and is therefore chosen for the intensity measurement. This is implemented by performing numerical integration of the the peak if it passes a programmable threshold voltage. 

The peak-to-valley distance scales with the longitudinal length of the bunch, and is relatively easy to measure. By fitting a second order polynomial to three sample points around maximum (minimum), a more accurate determination of the peak (valley) location is achieved, enhancing the the resolution of this measurement considerably.

These three measurements can be done with relatively low computational costs and perform well for our purposes. For more detailed motivations on the chosen waveform features, see \cite{masterohm}.

\subsection{Bunch phase and BCID association}
When the individual signals have been processed and their extracted features saved in waveform descriptors, the arrival time of each bunch can be matched with a clock edge. The time difference between the bunch arrival time and the clock edge defines the phase of the bunch and the number of clock ticks since the orbit pulse determines the BCID. The BCID association can also be done by matching the identified bunches to a reference LHC fill pattern, which could be useful in the unlikely event that the orbit signal is unavailable.

\section{BPTX Viewer: visualization of monitoring data}
\begin{figure}[h!!]
	\centering
	\includegraphics[width=0.5\paperwidth]{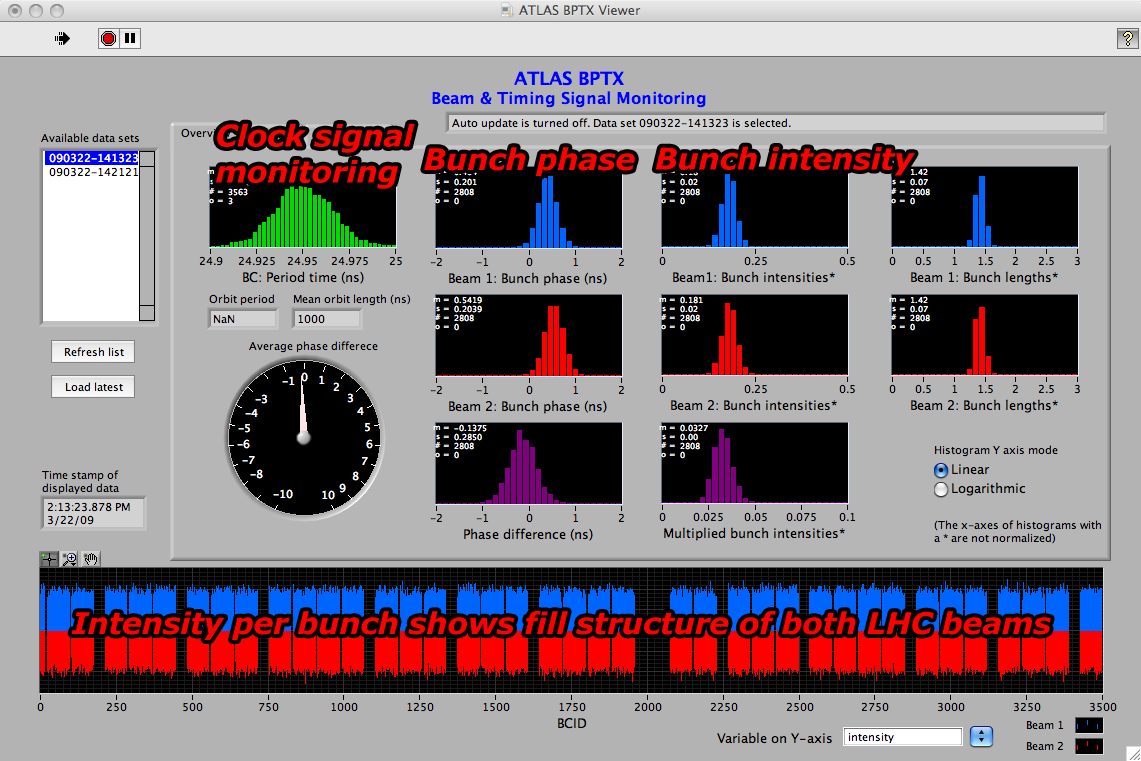}
	\caption{Screen shot of the BPTX Viewer application displaying measurements of simulated beam data.}
	\label{fig:bptxviewer1b}
\end{figure}

The objective of the BPTX monitoring system is to monitor the LHC beams and timing signals during ATLAS data taking. It is therefore important that the measurements are presented in a way that provides both overview and detail. The BPTX Viewer application provides an interactive graphical user interface for displaying monitoring data recorded with the BPTX Read-out program. Figure \ref{fig:bptxviewer1b} shows a typical monitoring display in BPTX Viewer. The panel on the left allows the user to browser through the data sets currently accessible on disk or to automatically display the most recent data. The bottom panel shows measured bunch quantities vs. BCID. In this case, the individual bunch intensity of beam 1 (blue) and beam 2 (red) is plotted for a simulated data set, and visualizes the fill structure and uniformity of both beams. The main panel shows summary histograms for the clock signal period time and distributions of the measured bunch phase, intensity and length for both beams.
\label{sec:bptxviewer}
\begin{figure}[h!!]
	\centering
	\includegraphics[width=0.5\paperwidth]{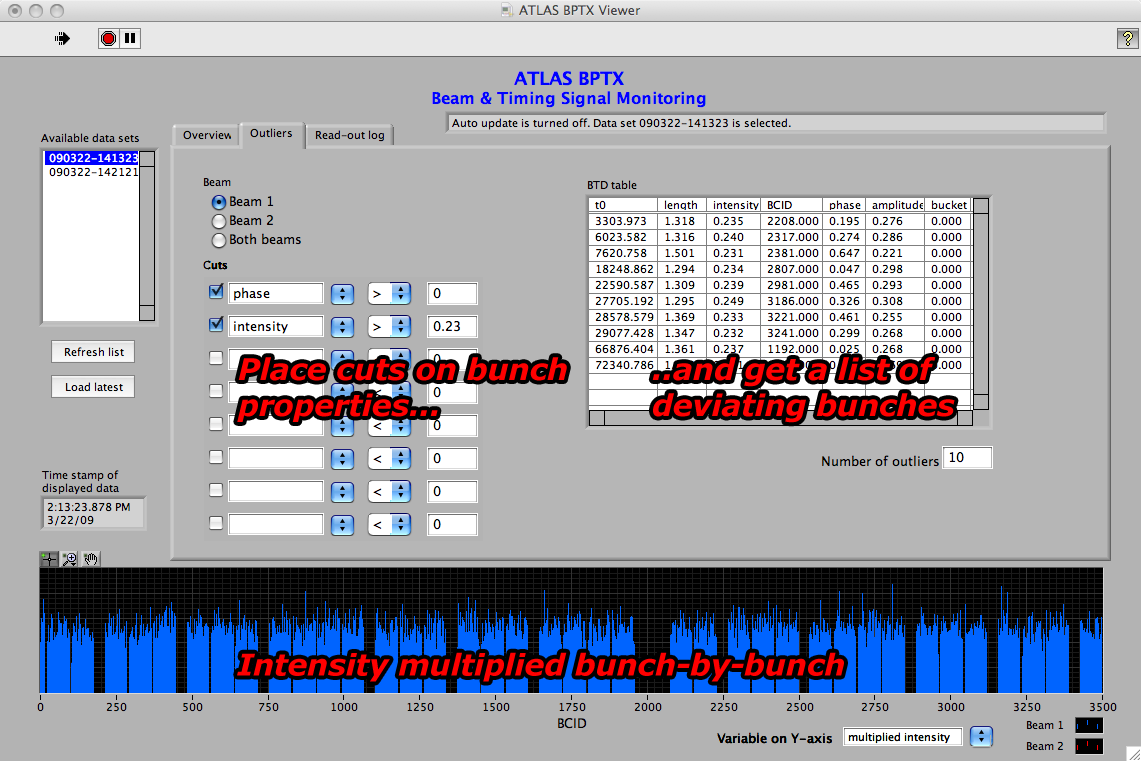}
	\caption{The Outliers tab in BPTX Viewer allows the shifter to identify deviating LHC bunches by placing cuts on their measured properties.}
	\label{fig:bptxviewer2b}
\end{figure}

The main panel has several tabs, and the second one, labeled ``Outliers'', offers an interface for finding deviating bunches. By defining a set of cuts on the measured bunch parameters, e.g. bunches with low intensity can be found easily and their information displayed in the table (see Figure \ref{fig:bptxviewer2b}). For example, displaying the bunches that have a phase that deviates from the mean value could reveal low-intensity, out-of-time bunches, so-called \emph{satellite bunches} or \emph{ghost bunches}. Although the signal model is yet to be tuned to data, simulations show that even two bunches in consecutive RF buckets\footnote{In the LHC, the RF frequency is 400\,MHz, resulting in a temporal distance of 2.5\,ns between the so-called RF buckets where the bunches can be situated.} can be distinguished and identified correctly by the software as long as they have comparable intensity. 

The third tab displays the current log file of the BPTX Read-out application for easy inspection in case the control room shifter suspects there are problems in the read-out or signal processing in the BPTX monitoring system.

\section{BPTX Waveform Viewer: a waveform display tool}
\label{sec:bptxwaveformviewer}
\begin{figure}[h!!]
	\centering
	\includegraphics[width=0.55\paperwidth]{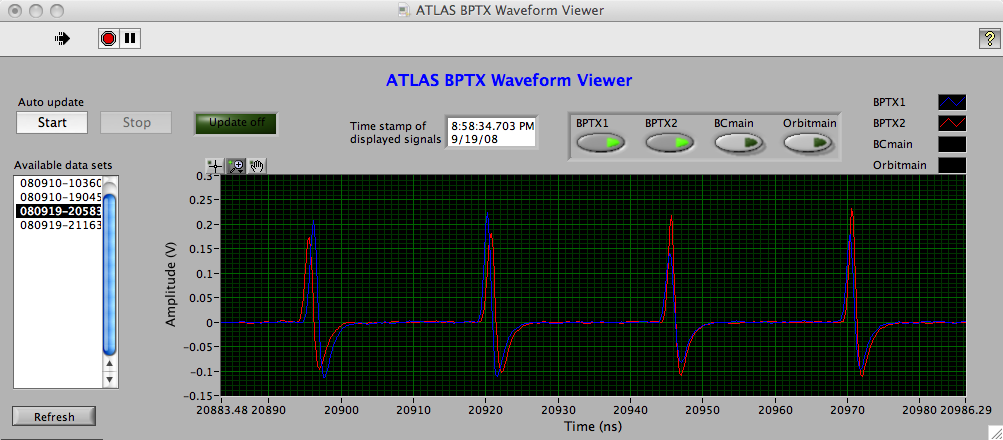}
	\caption{A screen shot from BPTX Waveform Viewer showing simulated BPTX signal waveforms.}
	\label{fig:bptxwaveformviewer}
\end{figure}

If the shifter in the ATLAS control room suspects that there is a problem with the timing signals or one of the beams, the waveforms available on disk can be inspected with the BPTX Waveform Viewer. Figure \ref{fig:bptxwaveformviewer} shows an example view of simulated BPTX signals for both LHC beams.

\section{Integration with the ATLAS online infrastructure and permanent storage}
The BPTX Publisher application runs in parallel to the BPTX Read-out program on the rack-PC in USA15, and is responsible for shipping summary data from the BPTX monitoring system to the ATLAS \emph{Detector Control System} (DCS)\cite{dcs}. By publishing the data to DCS, it is automatically made available to other ATLAS online services and recorded in the conditions database for permanent storage.

\section{Results from the first LHC beams}
At 10:18 AM on September 10, 2008, the LHC successfully circulated the first proton bunch. During the following week the BPTX system was used extensively to time in the ATLAS trigger system, and the BPTX monitoring system gave real-time information about the LHC beams to the shifters in the ATLAS control room. This section contains some plots from the monitoring system and presents a few preliminary results that indicate the resolution of its measurements.

\subsection{First LHC bunches in ATLAS}
\begin{figure}[h!!]
	\centering
	\includegraphics[width=0.5\paperwidth]{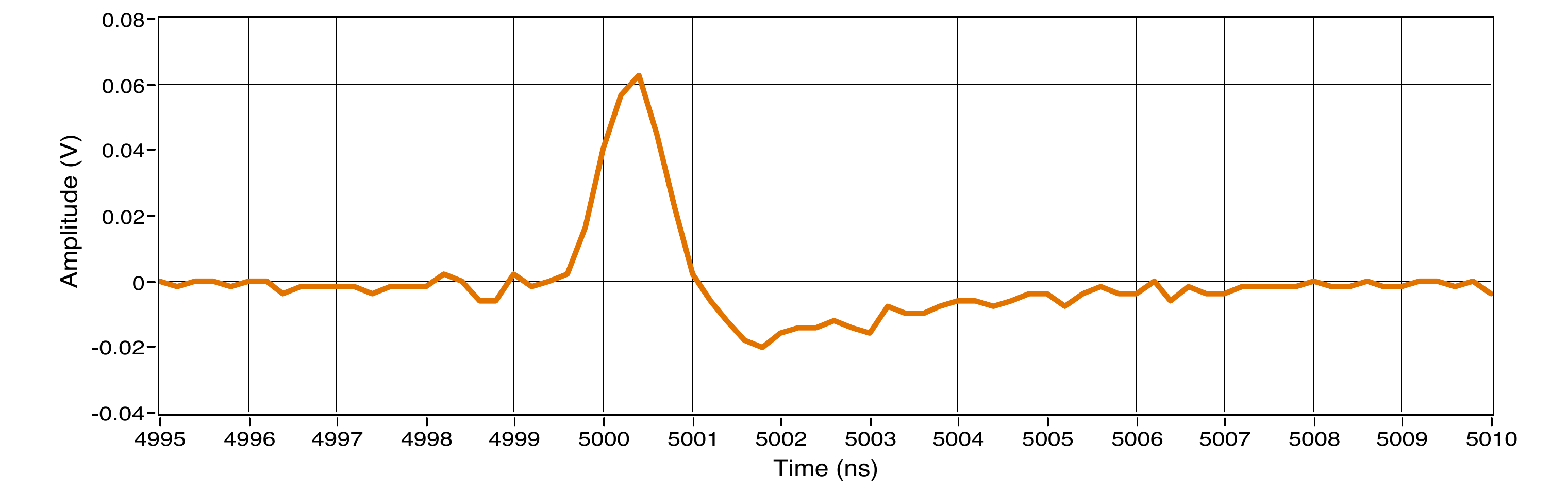}
	\includegraphics[width=0.5\paperwidth]{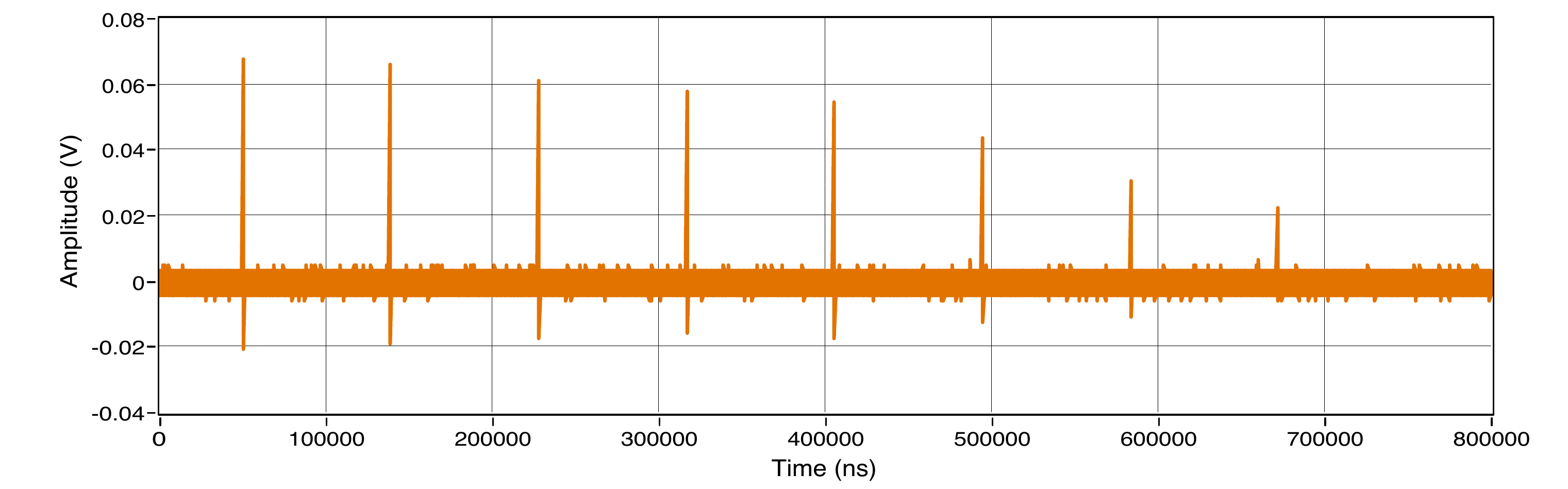}
	\caption{Top: The very first proton bunch approaching ATLAS as seen by the BPTX monitoring system on September 10, 2008. Bottom: A proton bunch recorded when circulating eight turns around the LHC.}
	\label{fig:bptxfirstbunch}
\end{figure}

Figure~\ref{fig:bptxfirstbunch} shows the first low-intensity LHC bunch on its way to ATLAS, as seen by the BPTX station 175\,m in front of ATLAS, before it hit the collimators and produced the first so-called \emph{splash event}. The amplitude of the pulse is consistent with a reported bunch intensity of $2\times10^{8}$ protons per bunch. For the nominal bunch intensity of $1.15\times10^{11}$ protons per bunch, the pulse amplitude is expected to increase by a factor of $\sim$500.

A few hours later, the LHC managed to circulate a single bunch eight turns around the LHC. Since the beam was not yet captured by the RF system, the pulse amplitude was reduced from turn to turn as can be seen in the bottom plot of Figure \ref{fig:bptxfirstbunch}. This can be explained by the debunching that is expected when the beam is not constrained longitudinally by an electric field.

\subsection{Longer injection}
\label{sec:longerinjection}
\begin{figure}[h!!]
	\centering
	\includegraphics[width=0.4\paperwidth]{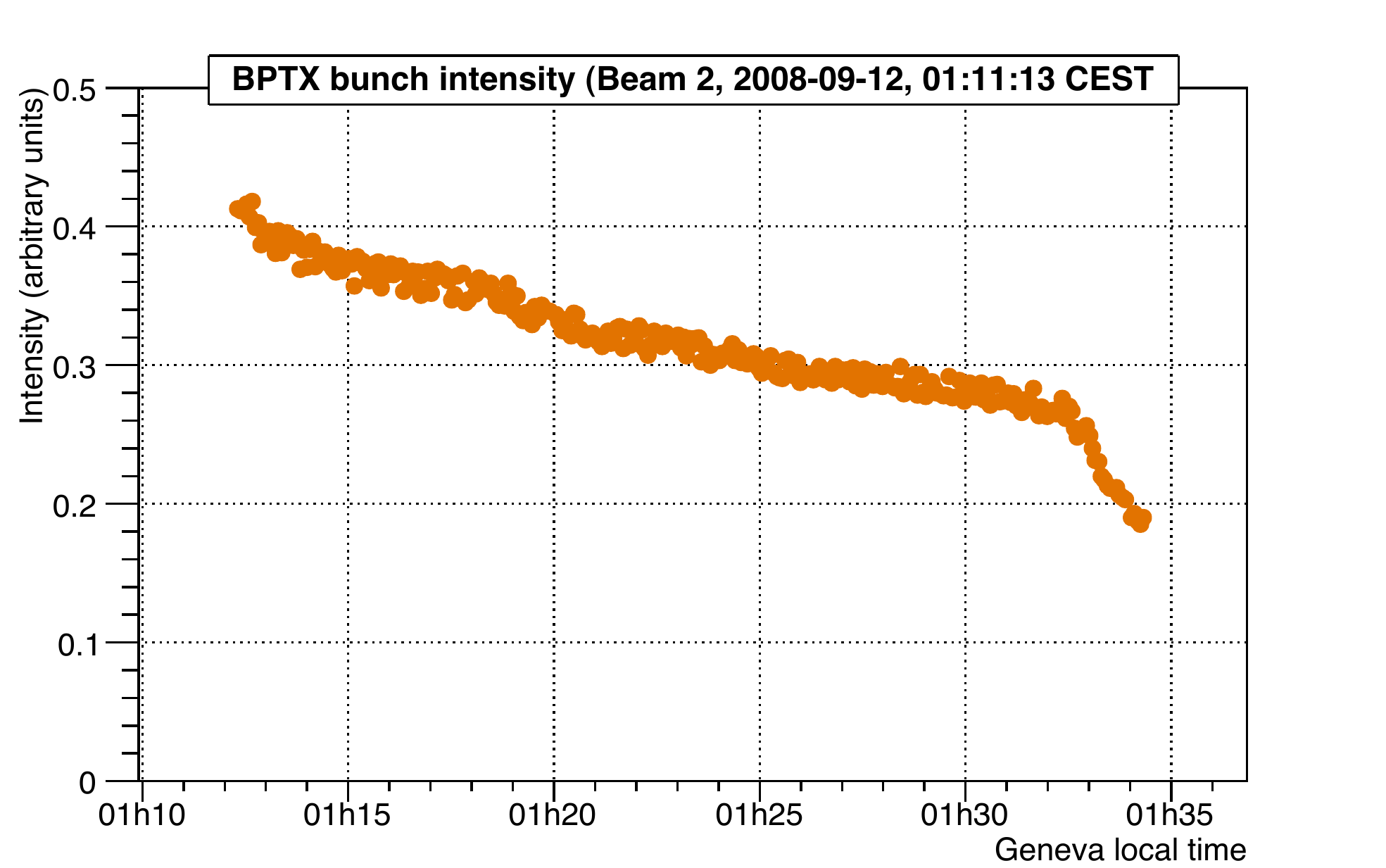}
	\caption{Beam intensity varying over a $\sim$20 min fill as measured by the BPTX monitoring system.}
	\label{fig:bptxlongerfill}
\end{figure}

A few days later, the LHC managed to capture a bunch with its RF system and circulate it for a longer period. Figure \ref{fig:bptxlongerfill} shows the beam intensity measured over time by the BPTX monitoring system for the longest LHC fill so far, with a total duration of about 20 minutes. The scatter of data points indicates that the resolution of this measurement is around 10 percent.

\begin{figure}[h!!]
	\centering
	\includegraphics[width=0.5\paperwidth]{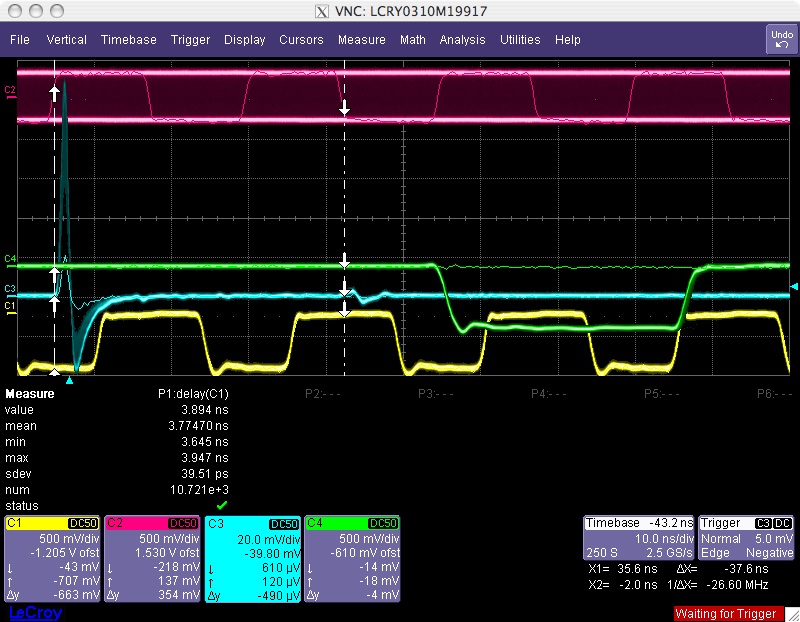}
	\caption{Oscilloscope trace taken in persistency mode during a long LHC fill.}
	\label{fig:bptxlongcoast}
\end{figure}

\subsection{Bunch phase resolution}
During the fill mentioned in Section \ref{sec:longerinjection}, the phase between the bunch and the closest edge of the clock signal was measured with a precision of 40\,ps.
This number includes not only the accuracy of the phase measurement, but also the clock jitter and possibly varying bunch arrival times. Figure~\ref{fig:bptxlongcoast} illustrates this phase stability with a picture from an oscilloscope that was running in persistency mode. Since all measurements were performed on the same bunch, potential effects on the arrival time determination due to varying bunch length are not included. 

\subsection{BPTX signal noise}
The period of single beam running offered the first possibility to measure the noise of the BPTX signals during operation with beam. By analyzing waveforms captured during this time, the standard deviation of the sample values in regions without bunch pulses was measured to be 1.7\,mV. On September 10, the LHC reported a bunch intensity of $2\times10^9$ protons, for which the measured BPTX pulse amplitude was 0.06\,V. Since the pulse amplitude varies linearly with bunch intensity, setting the threshold level to $5 \sigma_{noise}=8.5$\,mV the lowest possible bunch intensity that the BPTX system is sensitive to would be $3\times10^8$, corresponding to 0.26\% of the nominal LHC bunch intensity. As can be seen in Figure~\ref{fig:bptxlongcoast}, a small reflection of the bunch pulse was discovered in the recorded BPTX waveforms. The reflection is caused by impedance variations in the components that combine the signals from the four pick-ups, and is likely to scale with pulse amplitude. This feature could limit the sensitivity to detect satellite bunches in LHC fills with high bunch intensities, however there are plans to implement an appropriate filter to correct for it at the signal processing level.



\section{Conclusions}
The BPTX monitoring system in ATLAS is designed to monitor the relation between the LHC beams and timing signals to ensure the quality of the recorded event data. In addition, the structure of the beams and properties of the individual bunches can be measured and displayed with the BPTX application suite. During the first period of LHC running the BPTX signals were successfully used in the trigger system and the monitoring system measured both bunch intensity and phase with satisfactory performance, however, further calibration studies are needed to reach optimal performance.

\ack
The authors would like to thank the ATLAS Collaboration and its Level-1 Central Trigger group in which most of this work was carried out. We would also like to express our gratitude to the LHC machine groups.

\end{document}